# Expression for SU(4) multiplet and Masses of undiscovered Baryons in Standard Model

**Imran Khan**

*Department of Physics, University of Science and Technology,*

*Bannu 28100, Pakistan*

Email: immarwat@yahoo.com

*Abstract*—In particle physics, study of the symmetry plays very important role in order to get useful information about the nature. The classification and arrangements of subatomic particles is also necessary to study particle physics. Particles which are building blocks of nature are quarks, gluons and leptons. Baryons and Mesons composed of quarks were arranged by Gell-Mann and Okubo in their well-known Eight-Fold way up to SU(3) symmetry. Standard model of particles is composed of these particles. Particles in SU(4) also make some multiplets. However all the baryons with spin $J^P = 3/2^+$ and $1/2^+$ in these multiplets have not been observed till date. We have studied properties of the multiplets having spin $J^P = 3/2^+$ in an early work. In this paper the SU(4) multiplets with spin $J^P = 1/2^+$ have been organized and studied in an easy way. As a result some clues about the masses and iso-spins of the unknown hyperons have been obtained. These approximations about the characteristics of the unidentified baryons have been recorded in this article. Mass formula for the baryons having spin $J^P = 1/2^+$ in SU(4) multiplets have been extracted.

**Introduction**

Baryons are composed of three quarks (*qqq*). The three flavors up *u*, down *d*, and strange *s*, imply an approximate flavor *SU*(3), which requires that baryons made of these quarks belong to the multiplets on the right side of the 'equation' $3 \otimes 3 \otimes 3 = 10_S \oplus 8_M \oplus 8_M \oplus 1_A$. Here the subscripts indicate symmetric, mixed symmetry, or anti-symmetric states under interchange of any two quarks [1]. These were classified and arranged by Gell-Mann and Okubo in De-Couplets (*J* = 3/2, *l* = 0) and Octets (*J* = 1/2, *l* = 0) with +1 parities [2–4]. The Gell-Mann / Okubo mass formula which relates the masses of members of the baryon octet is given by; [2–4]

$$2(m_N + m_\Xi) = 3m_\Lambda + m_\Sigma \qquad (1)$$

While mass formula for de-couplets consists of equal spacing between the rows. The spaces are equal to an average value −151MeV. $M_\Delta - M_{\Sigma^*} = M_{\Sigma^*} - M_{\Xi^*} = M_{\Xi^*} - M_\Omega$  (2)

Gell-Mann used this formula and predicted the mass of the $\Omega^-$ baryon in 1962, equal to $M_\Omega = 1685 MeV$ [2]. Whereas actual mass of the $\Omega^-$ hyperon is equal to 1672 MeV, observed in 1964 [4]. Their mass difference is only 0.72 %, or in other words it was 99% true guess.

Now let's move towards baryon types made from the combination of four quarks, i.e. up $u$, down $d$, strange $s$ and charm $c$. These belong to $SU(4)$ multiplets. The $SU(4)$ multiplets numerology is given by $4 \otimes 4 \otimes 4 = 20_S \oplus 20_M \oplus 20_M \oplus 4_A$ [5].

The twenty particles having spin 3/2 and even parity +1 forming one of the $SU(4)$ multiplets have been studied in reference [6]. Now we are interested to study the twenty particles having spin 1/2 and even parity +1 forming another $SU(4)$ multiplet. These particles are in their ground states, with $l = 0$. For simplicity, natural unit of the mass is used throughout this article that is MeV, instead of MeV/$c^2$.

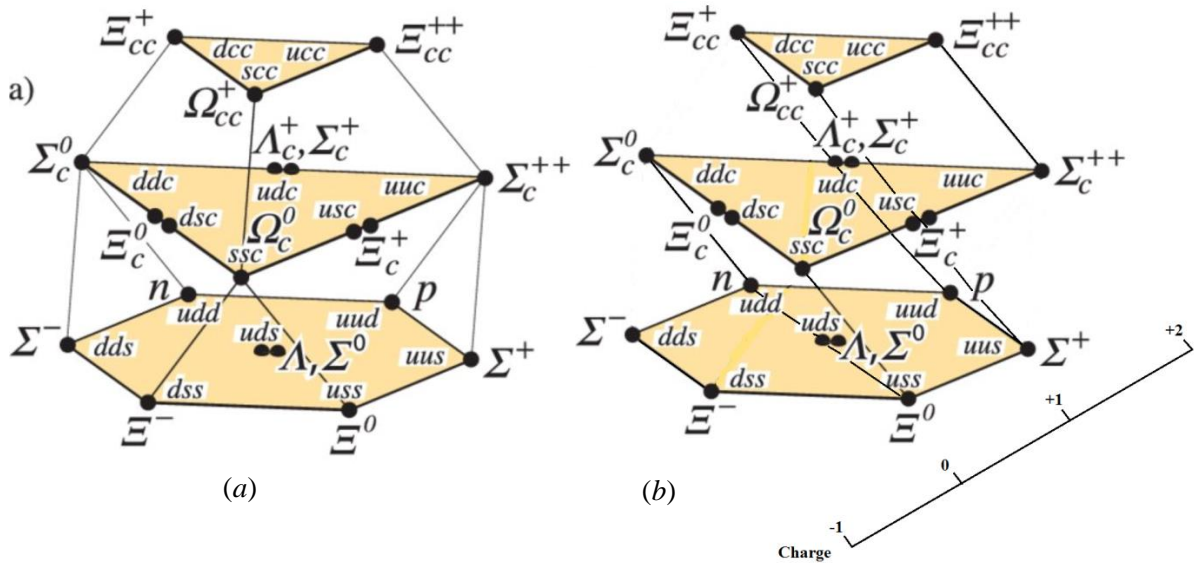

**Figure 1.** $SU(4)$ 20–plet of Baryons ($J^P=1/2^+$) made of $u$, $d$, $s$ and $c$ quarks, with an $SU(3)$ octet at the bottom [5] (a) $SU(4)$ 20–plet of Baryons ($J^P=1/2^+$) in another style with different layers (b).

In the $SU(3)$ framework the Gell-Mann / Okubo relation for the $J^P = 1/2^+$ Octets, equation (1) and the equal spacing rule for the $J^P = 3/2^+$ de-couplets, equation (2) work so nicely that we cannot abandon linear mass formulae for baryons [7]. Same behavior of the mass splitting of the particles may also be used to get expression for the particles having $J^P = 1/2^+$ and forming the multiplet in SU(4) as shown in figure (1a). Figure (1a) can be viewed from another angle, as shown in figure (1b). It is distributed in four different layers with increasing charge number.

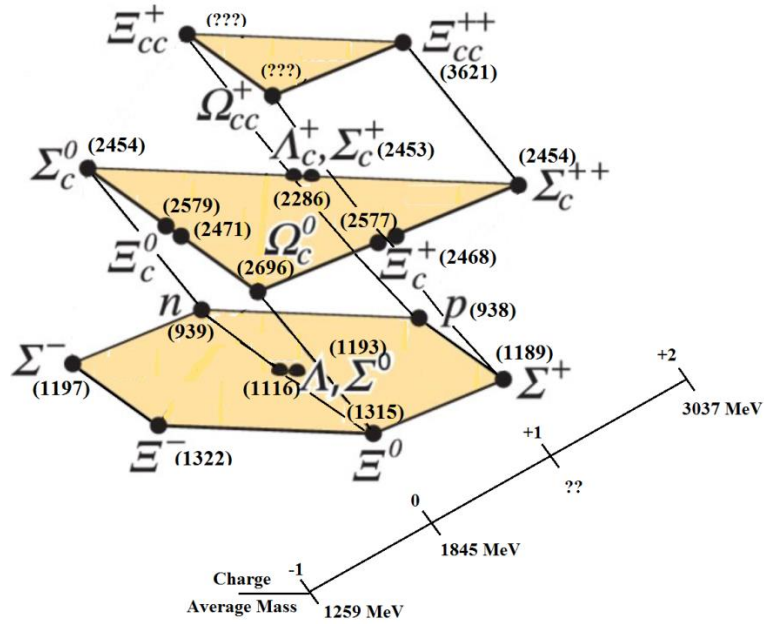

**Figure 2.** $SU(4)$ 20–plet of Baryons ($J^P=1/2^+$) made of $u$, $d$, $s$ and $c$ quarks, with different layers and their average masses.

In figure (2) quark contents of the particles are replaced with masses of the observed particles. Average of masses of particles in different layers with same charge number is written in front of each layer at the bottom right. Since $\Xi_{cc}^+$ and $\Omega_{cc}^+$ baryons are not discovered yet, therefore we cannot calculate average mass of the layer with charge number +1. However we can approximate masses of these two unknown particles with simple method, similar to the method used by Gell-Mann/ Okubo for SU(3) de-couplet. Average masses of the particles in layers with charge number -1 and 0, are given by $M_{-1} = 1259$ MeV and $M_0 = 1845$ MeV respectively. Average mass of the particles in layer with charge number +1, $M_{+1}$ cannot be obtained due to masses of two missing particles. Similarly average mass of the particles in layer with charge number +2, is given by $M_{+2} = 3037$ MeV.

Difference between average masses of the layers with charge number 0 and -1 is; $M_0 - M_{-1} = 586\,MeV$. Adding this value into the average mass of the layer with charge number

0, i.e. $M_0 = 1845$ MeV gives; $M_0 + 586\,MeV = 2431\,MeV = M_{+1}$. Similarly $M_{+1} + 586\,MeV = 3017\,MeV \approx M_{+2}$. Hence we can say that there is an equal spacing rule between these layers, given by;

$$M_2 - M_1 = M_1 - M_0 = M_0 - M_{-1} = 586\,MeV$$

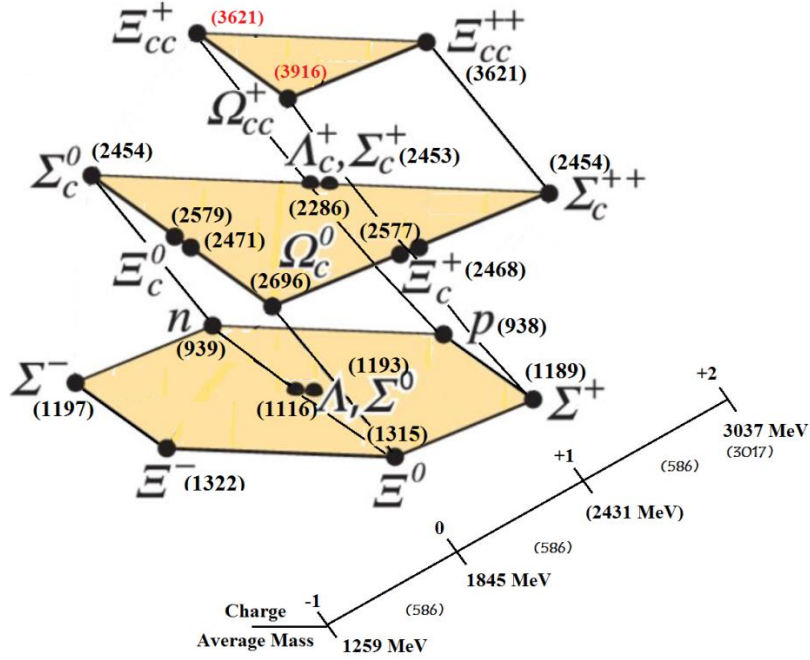

**Figure 3.** $SU(4)$ 20–plet of Baryons ($J^P=1/2^+$) made of $u$, $d$, $s$ and $c$ quarks, with different layers and their average masses and some calculations.

**Masses of the unknown particles**

As shown in figure (3), since layer with charge +1 has missing two particles; therefore we cannot calculate its average mass. Total mass of the six particles in this layer is equal to 11911 MeV. Average mass of the layer obtained using the above method is equal to 2431 MeV. Therefore, total mass of eight particles in the layer will be equal to 2431 × 8= 19448 MeV. Difference between calculated total mass of eight particles and six particles is equal to 19448 – 11911= 7537 MeV. Hence sum of the masses of two missing particles $\Xi_{cc}^+$ and $\Omega_{cc}^+$ will be approximately equal to 7537 MeV. But one particle $\Xi_{cc}^+$ will have the same mass as $\Xi_{cc}^{++}$ given by 3621 MeV, due to iso-spin symmetry, which is equal to $1/2\hbar$. Therefore mass of the $\Omega_{cc}^+$ particle will be equal to 7537 – 3621= 3916 MeV approximately. Or in more precise form we can take square-root of its value as uncertainty in the mass calculation, given by;
Mass of $\Omega_{cc}^+ = 3916 \pm 62$ MeV.

**Iso-Spin and its 3$^{rd}$ component of the particles**

In above paragraph it is stated that one particle $\Xi_{cc}^+$ will have the same mass as $\Xi_{cc}^{++}$ given by 3621 MeV, due to iso-spin symmetry, which is equal to 1/2ℏ. So we want to estimate iso-spin $I$ and its third component $I_3$. $SU(4)$ 20–plet of Baryons ($J^P=1/2^+$) shown in figure 1(a) may be presented in another style as shown in figure (4). Here iso-spin $I$ and its third component $I_3$ are presented in following figures as ($I$, $I_3$).

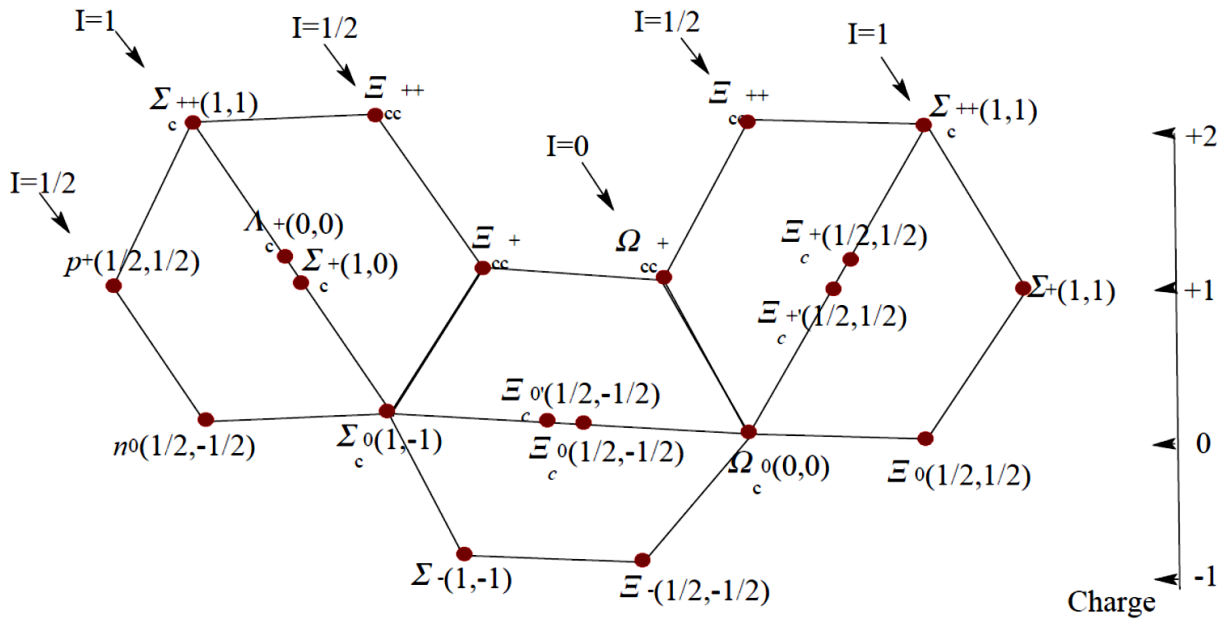

**Figure 4.** $SU(4)$ 20–plet of Baryons ($J^P=1/2^+$) in another style.

Here in this figure it may be observed that charge value in layers is increasing from bottom to top. Iso-spin has also a definite value in any diagonal layer (from top left side) from top to bottom. It is found from figure (4) that iso-spins of the undiscovered exotic hyperons $\Xi_{cc}^+$, $\Xi_{cc}^{++}$ and $\Omega_{cc}^+$ should be 1/2, 1/2 and Zero respectively. Similarly in order to find third component of iso-spin ($I_3$) of these particles, the figure (4) may be viewed from another angle, as shown in figure (5). It is observed that $I_3$ of particles in diagonal layers (from top right side) is decreasing by ½ for each layer. By setting the values of $I$ and $I_3$ for undiscovered hyperons according to the above procedure, it is found that $\Xi_{cc}^{+}$, $\Xi_{cc}^{++}$ and $\Omega_{cc}^{+}$ have values of $I_3$ equal to 1/2, -1/2 and Zero respectively.

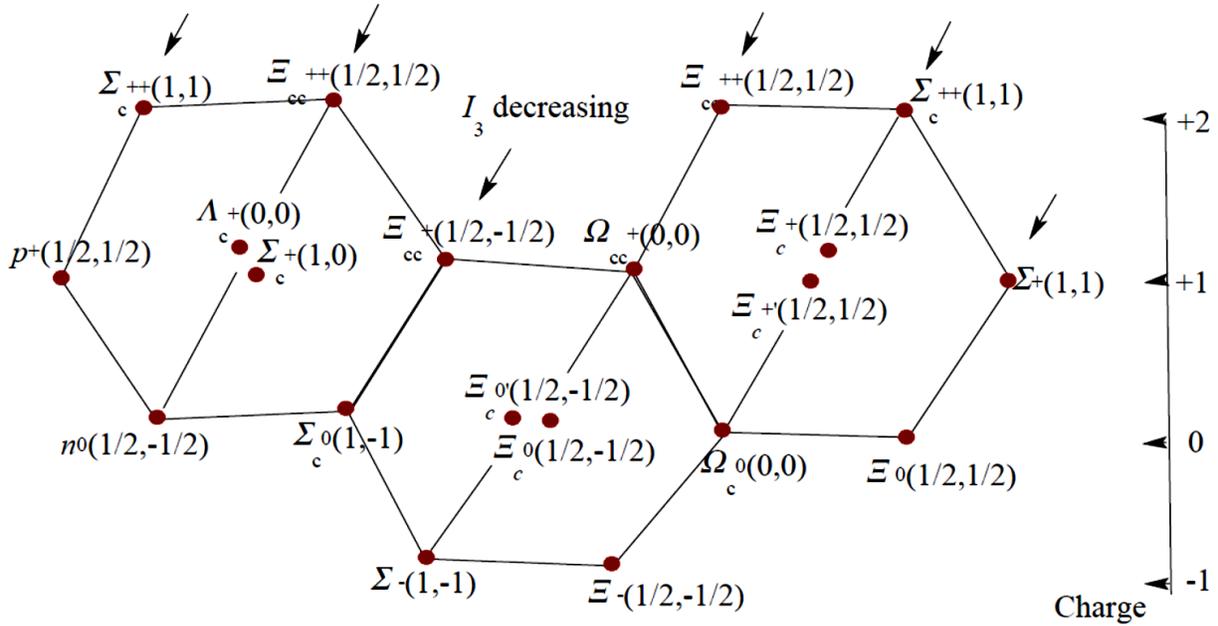

**Figure 5.** $SU(4)$ 20–plet of Baryons ($J^P=1/2^+$) for iso-spins calculations

**Conclusions**

In this article some characteristics of the baryons with $J^P = 1/2^+$ forming multiplets in SU(4) have been studied in an easy way. As a result some clues about the masses and other characteristics of the unknown baryons have been obtained. Mass splitting expression for the baryons having spin $J^P = 1/2^+$ in SU(4) multiplets have been obtained, given by; $M_2 - M_1 = M_1 - M_0 = M_0 - M_{-1} = 586\ MeV$. Masses of the missing particles $\Xi_{cc}^{+}$ and $\Omega_{cc}^{+}$ have been found to be approximately equal to 3621 ± 60 MeV and 3916± 62 MeV respectively. Iso-spin and third component of iso-spin of undiscovered hyperons $\Xi_{cc}^{+}$, $\Xi_{cc}^{++}$ and $\Omega_{cc}^{+}$ have been found using simple method. It is observed that Iso-spin and third component of iso-spin ($I$, $I_3$) of $\Xi_{cc}^{+}$, $\Xi_{cc}^{++}$ and $\Omega_{cc}^{+}$ particles should be (1/2, 1/2), (1/2,-1/2) and (0,0) respectively.